# Over half of the far-infrared background light comes from galaxies at $z \geq 1.2$

Mark J. Devlin[1], Peter A. R. Ade[2], Itziar Aretxaga[3], James J. Bock[4], Edward L. Chapin[5], Matthew Griffin[2], Joshua O. Gundersen[6], Mark Halpern[5], Peter C. Hargrave[2], David H. Hughes[3], Jeff Klein[1], Gaelen Marsden[5], Peter G. Martin[7,8], Philip Mauskopf[2], Lorenzo Moncelsi[2], Calvin B. Netterfield[8,9], Henry Ngo[5], Luca Olmi[10,11], Enzo Pascale[2], Guillaume Patanchon[12], Marie Rex[1], Douglas Scott[5], Christopher Semisch[1], Nicholas Thomas[6], Matthew D. P. Truch[1], Carole Tucker[2], Gregory S. Tucker[13], Marco P. Viero[8] & Donald V. Wiebe[9]

[1]Department of Physics & Astronomy, University of Pennsylvania, 209 South 33rd Street, Philadelphia, Pennsylvania 19104, USA. [2]School of Physics & Astronomy, Cardiff University, 5 The Parade, Cardiff CF24 3AA, UK. [3]Instituto Nacional de Astrofísica Óptica y Electrónica, Aptdo. Postal 51 y 72000 Puebla, Mexico. [4]Jet Propulsion Laboratory, Pasadena, California 91109-8099, USA. [5]Department of Physics & Astronomy, University of British Columbia, 6224 Agricultural Road, Vancouver, British Columbia V6T 1Z1, Canada. [6]Department of Physics, University of Miami, 1320 Campo Sano Drive, Coral Gables, Florida 33146, USA. [7]Canadian Institute for Theoretical Astrophysics, University of Toronto, 60 St George Street, Toronto, Ontario M5S 3H8, Canada. [8]Department of Astronomy & Astrophysics, University of Toronto, 50 St George Street, Toronto, Ontario M5S 3H4, Canada. [9]Department of Physics, University of Toronto, 60 St George Street, Toronto, Ontario M5S 1A7, Canada. [10]University of Puerto Rico, Rio Piedras Campus, Physics Department, Box 23343, UPR Station, Puerto Rico 00931. [11]INAF – Osservatorio Astrofisico di Arcetri, Largo E. Fermi 5, I-50125 Firenze, Italy. [12] Université Paris Diderot, Laboratoire APC, 10 rue Alice Domon et Léonie Duquet, 75205 Paris, France. [13]Department of Physics, Brown University, 182 Hope Street, Providence, Rhode Island 02912, USA.

**Submillimetre surveys during the past decade have discovered a population of luminous, high-redshift, dusty starburst galaxies[1–8]. In the redshift range $1 \leq z \leq 4$, these massive submillimetre galaxies go through a phase characterized by optically obscured star formation at rates several hundred times that in the local Universe. Half of the starlight from this highly energetic process is absorbed and thermally re-radiated by clouds of dust at temperatures near 30 K with spectral energy distributions peaking at 100 μm in the rest frame[9]. At $1 \leq z \leq 4$, the peak is redshifted to wavelengths between 200 and 500 μm. The cumulative effect of these galaxies is to yield extragalactic optical and far-infrared backgrounds with approximately equal energy densities. Since the initial detection of the far-infrared**

**background (FIRB)**[10]**, higher-resolution experiments have sought to decompose this integrated radiation into the contributions from individual galaxies. Here we report the results of an extragalactic survey at 250, 350 and 500 μm. Combining our results at 500 μm with those at 24 μm, we determine that all of the FIRB comes from individual galaxies, with galaxies at $z \geq 1.2$ accounting for 70 per cent of it. As expected, at the longest wavelengths the signal is dominated by ultraluminous galaxies at $z > 1$.**

The Balloon-borne Large-Aperture Submillimeter Telescope (BLAST) was designed as a precursor to the Spectral and Photometric Imaging Receiver (SPIRE) instrument[11,12] on the European Space Agency's Herschel Space Observatory. BLAST samples the emission from galaxies at 250, 350 and 500 μm close to the peak of the FIRB. The multiwavelength observations of BLAST fill the wavelength gap between the Multiband Imaging Photometer on NASA's Spitzer satellite (MIPS; 24–160 μm) and the windows at 850 μm and 1.2 mm accessible from the ground, providing a more complete understanding of galaxy evolution at this formative epoch.

The BLAST cosmological survey of the Great Observatories Origins Deep Survey[13] (GOODS-South) region combines a wide-area map of 8.7 square degrees with a deeper, confusion-limited map of 0.8 square degrees. These nested surveys, which are the focus of this paper, are hereafter known as the BGS-Wide and BGS-Deep observations, respectively. The depth of the BGS-Deep map was chosen to produce maps that have high signal-to-noise ratios and in which the fluctuations are dominated by pixel-to-pixel fluctuations in signals from galaxies rather than detector noise. The BLAST map (Fig. 1) overlaps some of the deepest multiwavelength data that exist in a cosmological survey, including radio, infrared, optical (Hubble Ultra Deep Field ) and X-ray (Extended Chandra Deep Field-South) surveys[13,14]. The BGS-Wide map was designed to match the coverage area and sensitivity of the near-infrared and mid-infrared Spitzer Wide-area Infrared Extragalactic survey[15]. By considering both of the data sets together, we derive a catalogue of sources that covers a large dynamic range, (a factor of 50 in flux density) with sufficient sensitivity to resolve the FIRB into individual galaxies. The area is large enough that the source counts are minimally affected by clustering. The BLAST survey contains approximately 500 sources with significant (>5$\sigma$) detections.

The brightness distribution, or number counts, of submillimetre sources probes the luminosity function in relation to redshift and can be used to constrain models for the formation and evolution of dusty, star-forming galaxies. Models that simultaneously fit the entire range of existing data (24–850 µm) include at least two distinct galaxy populations with different spectral energy distributions and evolutionary histories[16,17]. The BLAST data uniquely bridge these wavelengths across the energetic peak in the FIRB, and provide new strong constraints on the details of the evolution of these populations.

The number counts cannot be obtained directly from the distribution of detected sources in the BLAST catalogues because of several well-known biases. (1) The steep decrease in the counts with increasing flux density results in an Eddington bias whereby many more faint sources are boosted, rather than diminished in brightness, by noise fluctuations[18]. (2) The completeness of the survey falls off steeply at the faint end of the catalogue. (3) Multiple faint sources blend into brighter sources owing to a high surface density of the weakest sources and the 30–60″ resolution of the BLAST experiment. Rather than attempting to correct our source list for each of these effects, we estimate the counts from the distribution of pixel brightnesses in the entire map. This $P(D)$ analysis[19] implicitly handles all of the effects mentioned above, yet uses more of the information available to us than just the brightest pixels of the extracted point sources in our catalogues.

Figure 2 shows the BLAST number counts with the results from experiments spanning 24–850 µm, for comparison. The BLAST counts are consistent with a single power law at low flux density. We can place limits on the shape of the number counts at the faint end by comparison with the FIRB. Using the naive model that the power law continues down to some limiting flux $S_0$, with no sources fainter than $S_0$, we find that the inferred integrated surface brightness equals the FIRB when $S_0$ = 7.0 ± 1.3, 7.2 ± 1.7 and 4.6 ± 1.2 mJy at 250, 350 and 500 µm, respectively. A break in the counts, to a shallower slope, must occur at flux densities higher than the above limits to ensure that the integrated signal from the BLAST surveys is consistent with the amplitude of the FIRB. The increased sensitivity and higher resolution of SPIRE[11] should enable deeper surveys to detect the location of this spectral break.

BLAST is insensitive to individual sources significantly fainter than the 'confusion limit', which is conventionally defined as the flux at which sources reach a surface density of one object per 40 beams[19]. The confusion limits for BLAST and estimates of those for SPIRE based on the BLAST number counts are given in Table 1. We note that, through the $P(D)$ analysis, BLAST is sensitive to the number counts well below the confusion limit.

A stacking analysis using known positions of sources selected at different wavelengths is a powerful tool to estimate the contribution of a class of objects to the background at the wavelength of a particular map. The average brightness of these objects is calculated as the mean flux density measured at all of the positions in the map, a procedure that detects signals well below the threshold in signal-to-noise ratio used to make the BLAST catalogue. The contribution to the background from any stacked population is the product of its mean flux density and its surface density. A more detailed description of this method is provided in Supplementary Information.

We stack the 24-μm Far-Infrared Deep Extragalactic Legacy (FIDEL) survey catalogue (M. E. Dickinson *et al.*, in preparation) against the BLAST maps. The FIDEL catalogue contains sources brighter than 13 μJy. The stacked flux values are given in Table 2. We include a ~7–10% correction for completeness of the FIDEL catalogue based on extrapolation of 24-μm counts[20]. From the results of these calculations, shown in Table 2, we conclude that the FIDEL sources produce 75–100% of the FIRB as measured using the Far-Infrared Absolute Spectrophotometer (FIRAS)[21] on NASA's Cosmic Background Explorer.

Results similar to those from BLAST have been reported by stacking 24-μm Spitzer sources against the 70-μm and 160-μm maps[22]. Those data, recorded at wavelengths shorter than the peak intensity of the spectral energy distribution, are biased towards the low-redshift ($z \leq 1$) population that contributes to the FIRB. Surveys at 850 μm using the Submillimetre Common-User Bolometer Array (SCUBA) at the James

Clerk Maxwell Telescope (Mauna Kea, Hawaii) sampled the higher-redshift contribution but resolved only 20–30% of the FIRB above the confusion limit[5]. Techniques using gravitational lensing to identify fainter sources below the confusion limits of ground-based submillimetre telescopes have been claimed to resolve a larger fraction of the FIRB[23]. The wide spectral coverage of the BLAST data allows us to probe the Universe at both low redshift and high redshift. To demonstrate this, we use the Spitzer Infrared Array Camera (IRAC) colours to broadly divide the FIDEL catalogue into different populations to examine their relative contributions to the background at BLAST wavelengths. Using sources of known redshift in the GOODS Multiwavelength Southern Infrared Catalogue[24], we were able to define a simple cut in IRAC colours that effectively divides the catalogue into two galaxy populations above and below $z = 1.2$. At 250 μm, we find that there are approximately equal contributions to the background from sources above and below $z = 1.2$. We also note a systematic increase to 55% at 350 μm and 70% at 500 μm in the proportion of the background produced by sources at $z > 1.2$. These results are consistent with our expectation that objects in the longer-wavelength BLAST channels are dominated by the ultraluminous $z > 1$ galaxies discovered by SCUBA at 850 μm.

BLAST is the first instrument to make confusion-limited maps of the sky at wavelengths near the peak of the FIRB with enough sensitivity, sky coverage and angular resolution to identify a large number of sources, determine the detailed shape of the source counts and show that most of the FIRB comes from submillimetre sources already identified in deep 24-μm surveys. Combining this data with the results at 70 and 160 μm, we can account for essentially all of the FIRB with known galaxies. Combining the BLAST and future Herschel data at 250, 350 and 500 μm with spectroscopic and photometric redshifts will allow a determination of the optically obscured star-formation history of the Universe. These data will also provide catalogues of luminous submillimetre sources that will benefit from high-resolution imaging and spectroscopy from the Atacama Large Millimetre/submillimetre Array as we improve our understanding of galaxy formation and evolution in the high-redshift Universe.

**Supplementary Information** is linked to the online version of the paper at www.nature.com/nature.

**Acknowledgements** We acknowledge the support of NASA, the US National Science Foundation Office of Polar Programs, the Canadian Space Agency, the Natural Sciences and Engineering Research Council of Canada and the UK Science and Technology Facilities Council. We are grateful to B. Magnelli for help with the FIDEL 24-μm data. This research was enabled by the WestGrid computing resources and the SIMBAD and NASA/IPAC databases. We thank the Columbia Scientific Balloon Facility, Ken Borek Air Ltd and the mountaineers of McMurdo Station for their work.

**Author Information** Reprints and permissions information is available at www.nature.com/reprints. Correspondence and requests for materials should be addressed to M.J.D. (devlin@physics.upenn.edu).

**Table 1 Source confusion limits determined by BLAST**

| Wavelength (μm) | BLAST r.m.s. flux density from sources (mJy) | BLAST confusion flux density (mJy) | SPIRE confusion flux density (mJy) |
|---|---|---|---|
| 250 | 18 | 33 ± 4 | 22 ± 2 |
| 350 | 13 | 30 ± 7 | 22 ± 4 |
| 500 | 12 | 27 ± 4 | 18 ± 5 |

The fluctuations in the BLAST maps result from a combination of instrument noise and signals from sources. For each of the BLAST bands listed in the first column, the second column gives the residual map fluctuations from sources after subtracting instrument noise. The third and fourth columns give the formal confusion limits for BLAST and for SPIRE based on BLAST $P(D)$ analysis and the respective beam sizes. The confusion limit is the source brightness, $S_{conf}$, for which there is one source for every 40 beams. r.m.s., root mean squared.

**Table 2 The results of the 24-μm stacked flux**

| Wavelength (μm) | Stacked flux (nW m$^{-2}$ sr$^{-1}$) | Corrected flux (nW m$^{-2}$ sr$^{-1}$) | FIRB (nW m$^{-2}$ sr$^{-1}$) |
|---|---|---|---|
| 250 | 7.9 ± 0.5 | 8.5 ± 0.6 | 10.4 ± 2.3 |
| 350 | 4.5 ± 0.3 | 4.8 ± 0.3 | 5.4 ± 1.6 |
| 500 | 2.0 ± 0.2 | 2.2 ± 0.2 | 2.4 ± 0.6 |

For each of the BLAST bands in the first column, the second and third columns give the stacked and completeness-corrected fluxes, respectively. The fourth column is the expected flux determined by FIRAS[21]. The precision of the astrometry is 5″. To verify the astrometry, the stacked flux was shown to drop off in all directions when the map is shifted with respect to the 24-μm catalogue. A similar stack was made against the 2-Ms Chandra Deep Field-South survey catalogue[25] which resolves ~80% of the soft and hard X-ray backgrounds[26]. We find that these sources account for about 14% of the FIRB. Because active galactic nuclei account for 55% of the sources in the Chandra catalogue[25], we conclude that they only comprise 7% of the FIRB, although there may be a small contribution from Compton-thick active galactic nuclei that is missed.

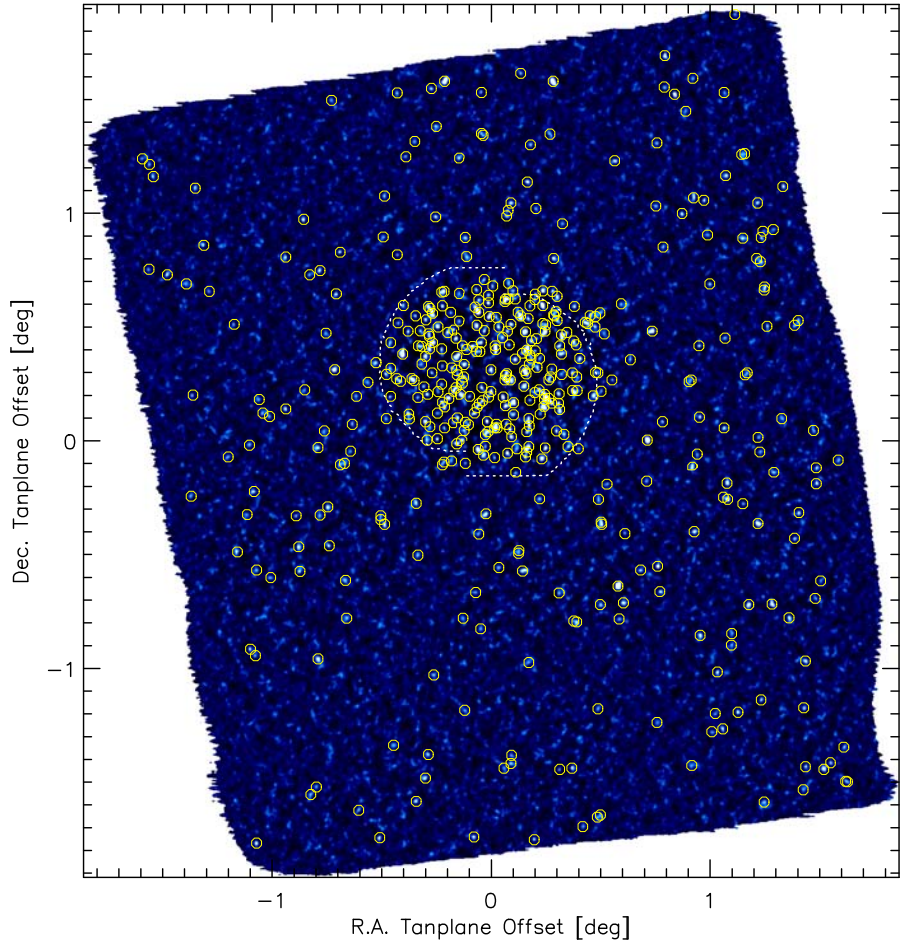

**Figure 1 Map showing the signal-to-noise ratio combination of all three BLAST bands for the entire BLAST GOODS-South observation.** The circles show the locations of the ~500 sources extracted from the map that are detected with a signal-to-noise ratio >5. The 0.8-square-degree BGS-Deep region, centred at right ascension (RA) 3 h 32 min 28 s, declination (dec.) –27° 48′ 30.00″ was mapped to $1\sigma$ depths of 11, 9 and 6 mJy at 250, 350 and 500 μm, respectively, and contains 253 sources. It is visible as the region of high density in detected sources in the map. The 8.7-square-degree BGS-Wide field was mapped to $1\sigma$ depths of 36, 31 and 20 mJy at 250, 350 and 500 μm, respectively, and contains 261 sources. A more detailed description of the map making is provided in Supplementary Information.

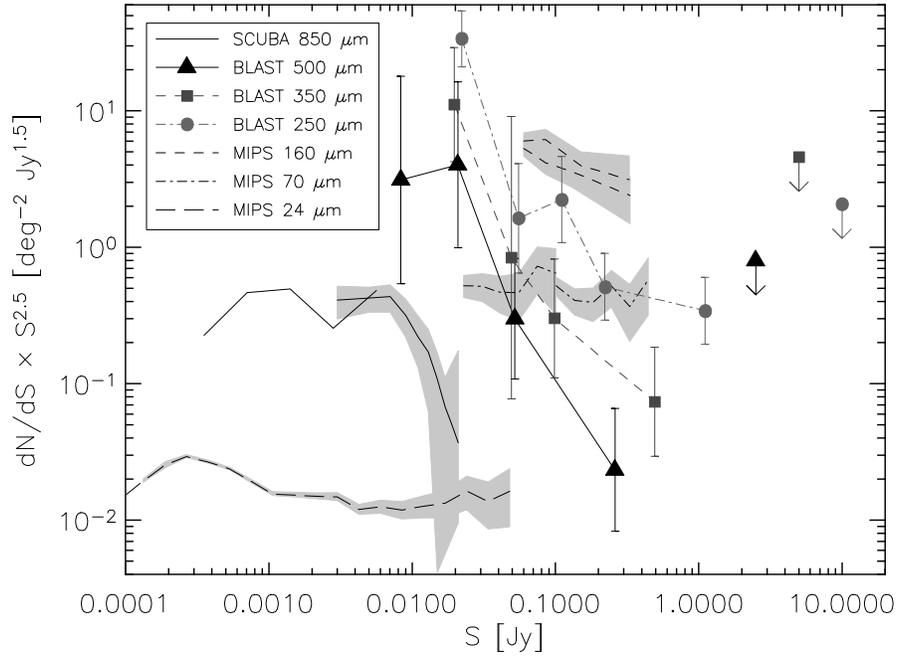

**Figure 2 The differential source counts at a variety of submillimetre and far-infrared wavelengths.** If there is no evolution in number density or luminosity, we expect number counts to be Euclidean with $dN/dS \propto S^{-2.5}$. We have therefore multiplied the counts by $S^{2.5}$ so that a Euclidean slope would be horizontal. The BLAST 250, 350 and 500-μm counts (circles, squares and triangles) are measured by $P(D)$ analysis, as described in the text and Supplementary Information. Other data shown (lines only) are SCUBA 850-μm counts[5,27] (solid lines); MIPS 160-μm counts[28] (short-dashed lines); MIPS 70-μm counts[28] (dash–dot lines); MIPS 24-μm counts[20] (long-dashed lines). The error bars are $1\sigma$. The shaded regions are $1\sigma$ error bands. The upper limits are $2\sigma$. Nearby sources, such as the brighter sources measured at 24 μm, appear Euclidean. The counts at 850 μm are steep, which is an indication of strong evolution. A striking feature of the BLAST counts is that the slope steepens with increasing wavelength: this is the transition from the short-wavelength part of the FIRB dominated by local sources to the longer-wavelength regime dominated by distant starburst galaxies. There appears to be a break in the 500-μm counts at the low end. However, the large error bar in the last bin and the fact that adjacent bins are inherently anti-correlated make the result not very significant.

Page 12 of 12

# Supplementary Information

## S1  Maps

The BLAST maps in this data release include two primary data products. The first set contains the "raw" maps produced by SANEPIC[29]. The SANEPIC algorithm was designed to provide the best estimate at *all* spatial scales in the image. The geometry of the BLAST instrument requires azimuthal scanning with a slow drift in elevation. When observations are made from extremely high latitudes near the South Pole, the azimuthal scans become essentially fast scans in right ascension, with a slow drift in declination. Varying detector baselines therefore result in an increase in large-scale noise along the declination direction.

The second set of maps suppresses large-scale noise to facilitate finding individual point sources. The 2-dimensional Fourier transform of the image is filtered to remove structure (primarily noise) on spatial scales larger than the size of the BLAST array projected on to the sky (approximately $14' \times 7'$). A similar technique has also been used for other recent bolometer experiments[30,31]. Since this filtering also slightly attenuates the signal of point sources, an essential part of this process is the production of an effective point spread function (PSF) that takes into account not only the telescope beam pattern, but also the effects of any additional filtering. Note that the signal and noise maps have units of MJy sr$^{-1}$, and the effective PSFs are in sr$^{-1}$, so that the flux density of a point source may be estimated as the peak value in the map, divided by the peak value of the corresponding PSF. Further details on the map data products, spectral band-passes and calibration are given in the accompanying *BLAST_README_06.txt*.

## S2  Source List

We provide source lists constructed independently for each BLAST band using a technique that is now standard practice in the (sub)millimetre extragalactic community. Using the filtered maps, we perform a noise-weighted convolution of the images with the effective PSFs. This operation is equivalent to calculating the maximum likelihood point-source flux density by which the PSF would have to be scaled to fit an isolated point source centred over each pixel in the map. We then produce source catalogues from the list of local maxima in the smoothed maps. Once each peak pixel has been identified, we improve the astrometry by fitting the vicinity of the peak with a circular Gaussian in which the position and amplitude can vary.

In the wide/shallow part of the maps, individual flux densities and positions should be fairly good estimates, since the surface density of sources is low. However, the deep part of the maps is confused and many faint sources could potentially blend to form single, apparently brighter sources. No attempt has been made to correct for this in the released catalogue—we simply quote flux densities, noises, and positions for all signal-to-noise ratio (S/N) peaks with significances greater than $3\sigma$. We do not presently provide completeness or flux bias corrections for this list. These issues are addressed in the next section.

## S3  Source Counts

Several well known effects hinder the process of obtaining un-biased estimates of the underlying differential source counts distribution, $dN/dS$ directly from the source catalogue: (i) extra apparently bright sources are produced at the expense of fainter sources that blend together in confused regions of the map; (ii) Eddington bias also produces greater numbers of brighter rather than fainter sources because of the steeply falling $dN/dS$; (iii) the "completeness" of the catalogue drops at fainter flux densities; and (iv) the number of spurious false positives increases at lower S/N thresholds. We experimented with attempts to correct for these effects using similar methods [5,31], but found this to be very dependent on the prior model of the counts in the confusion-dominated deep region of the BGS-Deep maps.

Instead of using our source catalogues to measure the source counts, we developed a "$P(D)$" analysis to fit directly an underlying model of $dN/dS$ to the maps themselves, implicitly handling all of the issues mentioned above [32]. To avoid biasing its shape, $dN/dS$ is parameterized by a list of nodes at fixed flux densities, $S$, with amplitudes that are allowed to vary. The surface density of sources between these nodes are then interpolated using power-laws. For a given model realization we calculate the expected signal distribution, the so-called "probability of deflection". We then calculate the likelihood that the true signal histogram could have been a realization of our model. We find values for the parameterized $dN/dS$ that maximize the likelihood of the data. The observed and best-fit signal histograms at $250\mu$m are shown in Figure S1 to illustrate this procedure. To characterize the uncertainties in the differential source counts we perform a Markov Chain Monte Carlo (MCMC) simulation to sample the likelihood function around the global maximum.

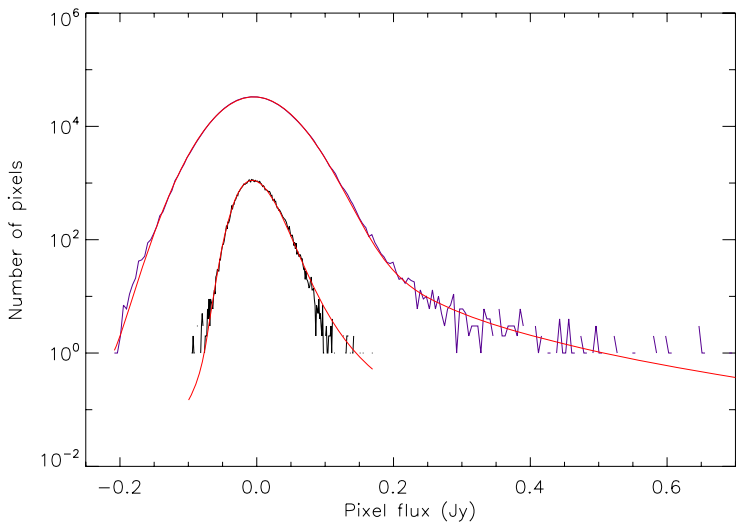

*Figure S1. The signal histogram derived from the 250 $\mu$ m map (black jagged line), compared with the expected histogram from the maximum likelihood model for $dN/dS$ (red line) derived from the $P(D)$ analysis. The top and bottom curves represent the wide/shallow and narrow/deep parts of the map, respectively. These are shown separately for illustrative purposes, but in practice the entire map is fit simultaneously. If the map were devoid of sources, an approximately symmetric Gaussian noise distribution would be expected in each case. The positive signal tail is caused by sources, including large amounts of signal below the confusion limit.*

The maximum-likelihood values for the real data, and independent 1σ uncertainties, for all of the bands, are given in Table S1. The Pearson Correlation matrices for the parameters inferred from the MCMC simulation are given in Tables S2 - S4. Note that adjacent nodes are highly anti-correlated: if the counts in a given range are higher, it may only be at the expense of removing sources from adjacent ranges. It is for this reason that the large drop in the 500 $\mu$m counts in the faintest bin is less significant than what is implied by Figure 2 in the main paper; if the true source counts at the next brightest node tend to pass through the lower error bar, the counts at the faintest node must increase by a large factor.

Table S1: Maximum-likelihood models in each band for the differential source counts, dN/dS, parameterized by the log of the surface density at fixed nodes, and with a power-law interpolation between the nodes assumed. Uncertainties are quoted as 1-$\sigma$ errors.

| 250 $\mu$m | | 350 $\mu$m | | 500 $\mu$m | |
|---|---|---|---|---|---|
| Node (Jy) | Value (log[deg$^{-2}$Jy$^{-1}$]) | Node (Jy) | Value (log[deg$^{-2}$Jy$^{-1}$]) | Node (Jy) | Value (log[deg$^{-2}$Jy$^{-1}$]) |
| 0.0001 | 4.6 ± 4.1 | 0.00005 | 9.6 ± 3.1 | 0.00003 | 10.5 ± 9.7 |
| 0.022 | 5.7 ± 0.2 | 0.02 | 5.3 ± 0.4 | 0.0084 | 5.7 ± 0.8 |
| 0.055 | 3.4 ± 0.4 | 0.049 | 3.2 ± 1.0 | 0.021 | 4.8 ± 0.6 |
| 0.11 | 2.7 ± 0.3 | 0.1 | 2.0 ± 0.4 | 0.052 | 2.7 ± 0.4 |
| 0.22 | 1.35 ± 0.25 | 0.49 | -0.36 ± 0.40 | 0.26 | -0.17 ± 0.45 |

The dN/dS curves estimated directly from source catalogues (number of sources in a flux density bin, divided by the bin width and area of the survey) are shown in Figures S2 - S4. The shapes of these curves are dominated by the effects of confusion, bias, and incompleteness in the source lists. To illustrate this we use 100 Monte Carlo simulations. Given the maximum likelihood model for the differential counts in each band, we first draw $N \pm \sqrt{N}$ source flux densities from the distribution at random, where $N$ is the total number of sources with flux densities > 0.01 mJy, a range chosen to encompass the lowest node of the parametric model (0.1 mJy). Each source is assigned a random position (unclustered), and represented by a $\delta$-function with an amplitude given by its flux density. This "point-source" map is then convolved with the telescope beam to give the noiseless representation of the sky that would be observed by the telescope. To this map we add a realization of noise with a spatially-varying rms that matches the real data, producing a data set with realistic signal and noise properties. A comparison of the actual catalogue counts, and counts expected from the underlying $dN/dS$ distribution (95% confidence interval of the simulations) are shown to have remarkable agreement. These "catalogue counts" in each band clearly exhibit two peaks, and only agree with the underlying $dN/dS$ (dashed lines) at the bright end. The bumps are produced by sources near the 4-σ detection thresholds in each of the BGS-Wide and BGS-Deep regions, where the effects of positive flux density biases are most prevalent. Leftwards of each of these peaks incompleteness dominates, causing the steep drops.

Table S2: Pearson Correlation matrix for the parameterized dN/dS model at 250$\mu$m.

| Node(Jy) | 0.00001 | 0.022 | 0.055 | 0.11 | 0.22 | 1.1 |
|---|---|---|---|---|---|---|
| 0.00001 | 1.00 | −0.92 | 0.48 | −0.084 | 0.097 | −0.090 |
| 0.022 | | 1.00 | −0.76 | 0.26 | −0.064 | 0.007 |
| 0.055 | | | 1.00 | −0.62 | 0.096 | 0.060 |
| 0.11 | | | | 1.00 | −0.580 | 0.240 |
| 0.22 | | | | | 1.000 | −0.720 |
| 1.1 | | | | | | 1.000 |

Table S3: Pearson Correlation matrix for the parameterized dN/dS model at 350$\mu$m.

| Node(Jy) | 0.00005 | 0.02 | 0.049 | 0.1 | 0.49 |
|---|---|---|---|---|---|
| 0.00005 | 1.00 | −0.98 | 0.82 | −0.19 | −0.046 |
| 0.02 | | 1.00 | −0.90 | 0.30 | −0.032 |
| 0.049 | | | 1.00 | −0.57 | 0.250 |
| 0.1 | | | | 1.00 | −0.750 |
| 0.49 | | | | | 1.000 |

*Table S4: Pearson Correlation matrix for the parameterized dN/dS model at 500μm.*

| Node(Jy) | 0.00003 | 0.0084 | 0.021 | 0.052 | 0.26 |
|---|---|---|---|---|---|
| 0.00003 | 1.00 | −0.95 | 0.78 | −0.44 | 0.23 |
| 0.0084 |  | 1.00 | −0.89 | 0.55 | −0.26 |
| 0.021 |  |  | 1.00 | −0.80 | 0.38 |
| 0.052 |  |  |  | 1.00 | −0.62 |
| 0.26 |  |  |  |  | 1.00 |

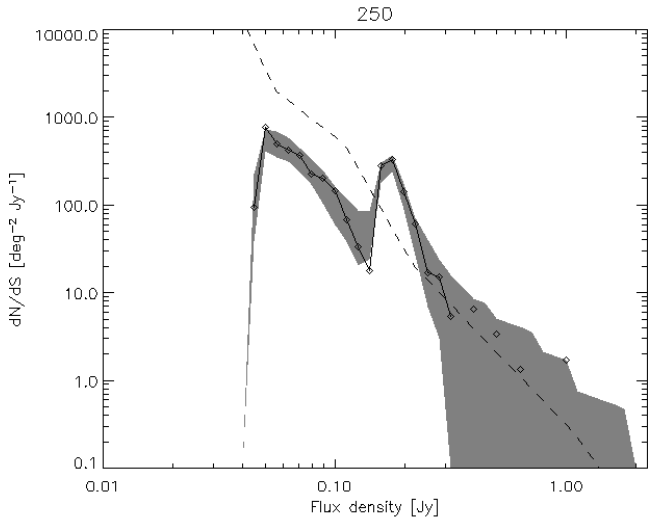

*Figure S2: Comparison of dN/dS with the catalogue at 250μm. The dashed line is the best-fit differential counts distribution from the P(D) analysis. The shaded region gives the 95% confidence interval of simulated 4-σ catalogue counts distributions from multiple realizations of the best-fit model, including noise and instrumental effects, and then counting sources in flux bins and dividing by the survey area—i.e., this is what we should measure in our survey if the dashed line is correct. The solid line gives the catalogue counts from the real data. The "spikes" at 0.04 Jy and 0.15 Jy result from the superposition of sources detected in the deep and wide survey areas respectively. The completeness drops to the left of the spikes. Spurious detections, Eddington bias, and confusion increase the counts at the locations of the spikes. These effects are decreased at higher flux densities, and hence higher S/N.*

## S4 Stacking

To calculate the contribution to the Far-Infrared Background (FIRB) produced by a given population of sources, we determine the average flux densities in the BLAST bands at the positions of all objects in flux density bins of a given catalogue. We then multiply by the surface density of sources in that flux density bin and sum over all the bins. This "stacking" formula is derived in Section S4.1. We have used a 24 $\mu$m catalogue from the *Spitzer* Far-Infrared Deep Extragalactic Legacy (FIDEL) survey, for which IRAC positions in the SIMPLE survey [33,34] have been used as a positional prior. The coverage of this catalogue with respect to BLAST is shown in Figure S5. The application of this catalogue to

BLAST stacks is described in Sections S4.2 - S4.3.

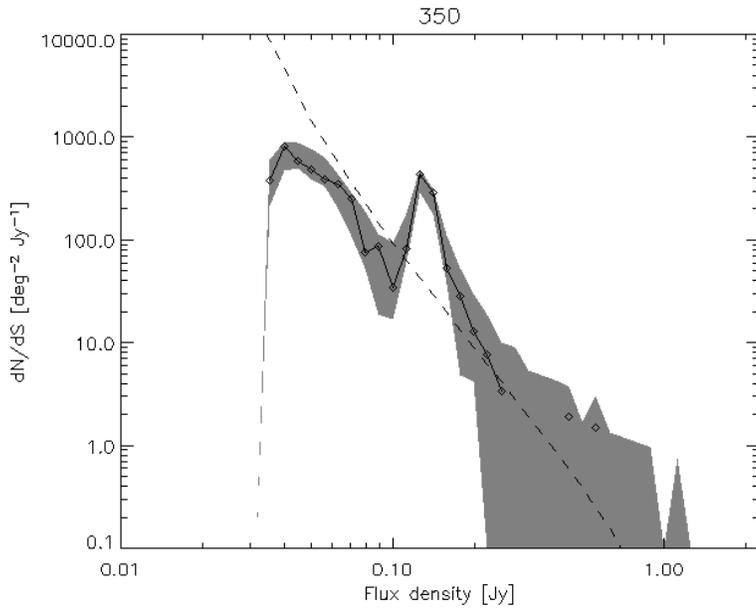

*Figure S3: Similar to Figure S2, this plot compares $dN/dS$ with the 4-$\sigma$ catalogue at 350$\mu m$. Again this shows that we obtain the expected counts in our catalogue given the underlying model (dashed line) and the various biases at low S/N.*

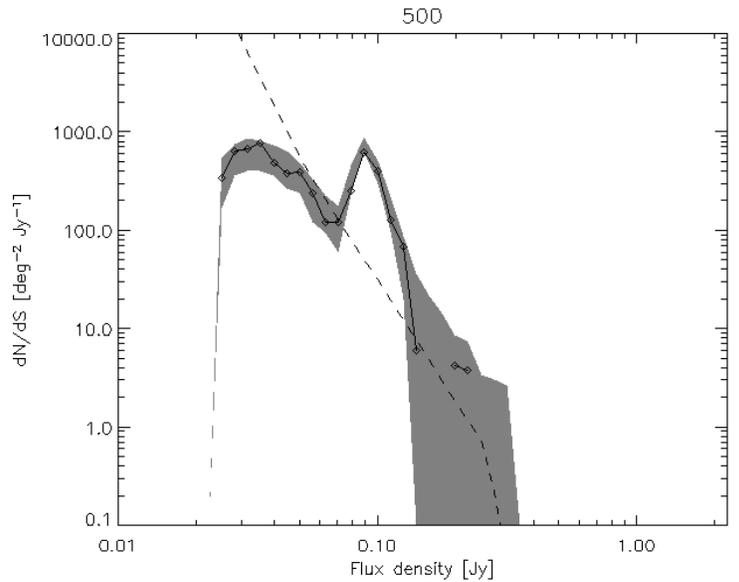

*Figure S4: Similar to Figure S2, this plot compares $dN/dS$ with the 4-$\sigma$ catalogue at 500$\mu m$. Once again the counts from the catalogue (solid line) match well the simulations (shaded region) given the underlying model (dashed line).*

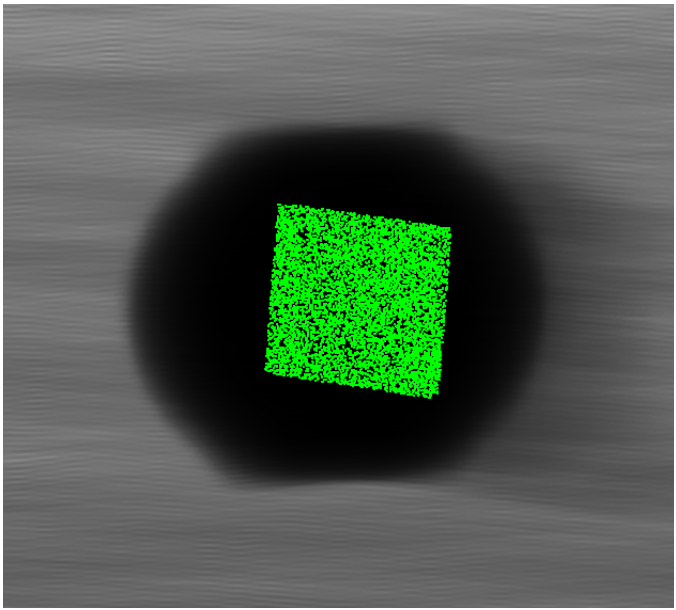

*Figure S5: Locations of FIDEL 24 $\mu m$ sources (green symbols) compared to the BLAST coverage (greyscale proportional to map noise variance). The dark circular region corresponds to the BGS-Deep BLAST coverage, with an area of 0.8 $deg^2$. The FIDEL catalogue covers an area of 0.2 $deg^2$ within this region. The mean value of the BLAST map across the field is zero, but the average stacked flux produced by the FIDEL sources is sufficient to explain the bulk of the FIRB measured by COBE.*

## S4.1 Development of the Stacking Formulae

We model the sky by a map of pixels indexed $j$, with flux densities $D_j$. Suppose we have a catalogue of sources from some other experiment, potentially at a different wavelength, and that the number of sources from this catalogue located within each pixel is $N_j$. Denote the mean value of $N_j$ with $\mu$. In all that follows we will assume that the sources in the catalogue are not correlated, such that $N_j$ is a random, Poisson distributed number. We will also ignore the effects of source clustering.

If items in the catalogue produce flux densities that are $S_0$ on average then, along with whatever else is in the sky, there will be a contribution of $S_j = S_0 N_j$ to each pixel. If a sky containing this signal were observed with BLAST the resulting map would be the convolution of $S_j$ with the PSF, and with the mean value subtracted.

We have already mapped the sky with BLAST, so we have the following problem: given a map and a catalogue, can we find the mean flux density per source, $S_0$, at the BLAST wavelengths? We approach this problem by considering our map, $D_j$, and our external catalogue distribution, $N_j$, as shapes on the sky, and searching for the amplitude, $S_0$, of $N_j$ that matches $D_j$. The cross-correlation of $D_j$ with $N_j$ does just this.

$$D_j \otimes N_j = \sum_j D_j N_j. \tag{1}$$

The expectation value of this sum is

$$\langle D_j \otimes N_j \rangle = S_0 \sum_j N_j^2 = S_0 \mathsf{N} \sigma_N^2, \tag{2}$$

where $\mathsf{N}$ is the total number of pixels and the variance of a Poisson distribution

$$\sigma_N^2 = \mu. \tag{3}$$

The net result is that the cross-correlation of a catalogue can be used to estimate the flux density per source.

An additional re-arrangement of Eq. 2 makes this result more useful. Notice that the sum runs over all pixels, with the weight of each pixel proportional to the number of catalogue sources found in it, and that zero weight is given to pixels that do not contain a source ($N_j = 0$). This can be written as a sum over catalogue entries with unit weight:

$$S_0 = \langle D_j \otimes N_j \rangle \mathsf{N} \mu = 1 \mathsf{N} \mu \sum_j D_j N_j = 1 n \sum_k D_k, \tag{4}$$

where $k$ is the index of the source catalogue, $D_k$ is the measured flux density in the map pixel that contains the $k$th catalogue entry and $n$ is the number of catalogue entries, $n = \mathsf{N}\mu$. This expression is the simple average flux density in the map over all positions in the source catalogue. It is perhaps counterintuitive that no additional correction is needed to account for cases in which the catalogue sources are highly confused. Note, however, that this statement is only strictly true in the absence of clustering, and if the instrumental noise is well-behaved.

To obtain uncertainties we repeat this calculation for a set of random locations in the map, and measure the sample standard deviation of the resulting stacks. This procedure accounts for uncertainties caused both by instrumental noise and confusion.

Dividing a source catalogue into flux density bins of width $\delta s_i$, we measure the source surface density, $N(s_i)$, and stacked values, $S_0(s_i)$, in each bin, to calculate the contribution of the catalogue sources to the background

$$I = \sum_i S_0(s_i) N(s_i) \delta s_i. \tag{5}$$

## S4.2 FIDEL Catalogue Completeness

To determine the completeness of the FIDEL catalogue we compare the density of sources with the published, completeness-corrected distribution[20] shown in Figure S6. The comparison indicates that the three lowest flux bins are 63%, 80% and 96% complete. The completeness corrections at the faint end add 7%, 8% and 11% to the total sky intensity summed over all bins at 250, 350 and 500 $\mu$m. The resulting stacking analysis shows that essentially all of the submillimetre background in the BLAST bands can be accounted for by sources identified at 24 $\mu$m by *Spitzer* using Eq. 5.

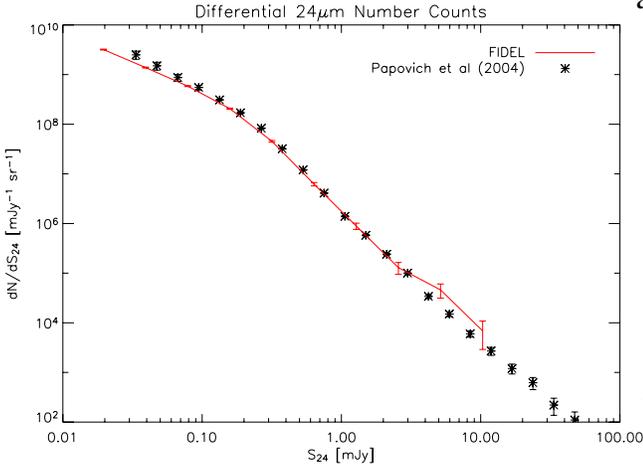

*Figure S6: Comparison between 24μm source counts in the FIDEL catalogue, and the completeness corrected counts [20]. We use this comparison to calculate completeness correction factors for the three lowest bins in the FIDEL counts.*

## S4.3 IRAC Colour Selection

One of our primary results is the measurement of relative fractions of sources above and below $z = 1.2$ that contribute to the FIRB in each BLAST band. We use *Spitzer* IRAC colours to broadly classify each object in the FIDEL 24 $\mu$m catalogue. To calibrate our classification scheme, we examine the MUSIC catalogue for which IRAC colours and redshifts are known. We then explored the IRAC colour space, following similar studies by other authors[35,36], and found that a simple cut effectively divides sources into populations of low- and high-$z$ sources as shown in Figures S7 and S8. Stacking these two sets separately yields the relative contributions to the FIRB.

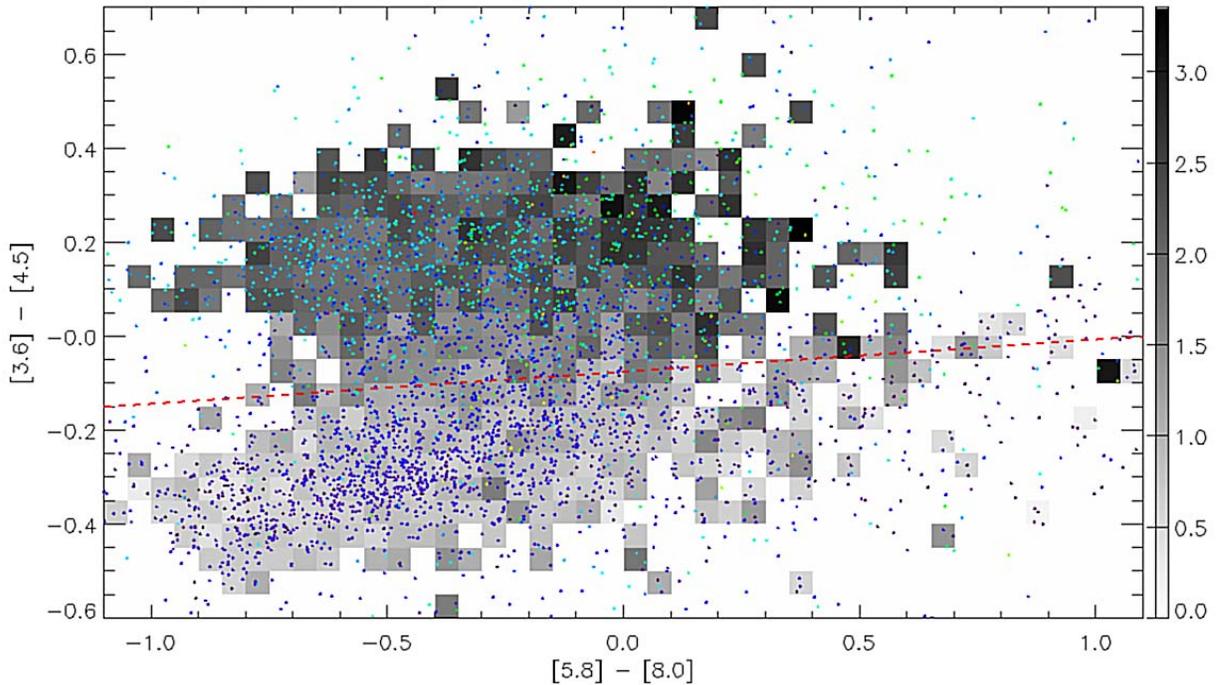

*Figure S7: IRAC AB magnitude colour-colour plot for the MUSIC and FIDEL catalogues. The grey intensity plot gives the average known redshifts (from white at $z = 0$ to black at $z = 3.5$) of MUSIC sources in bins that contain at least 5 objects. The symbols correspond to individual objects in the FIDEL catalogue, indicating that they span a similar range in this colour space (colours of the points represent redshift, going from purple for $z = 0$ through blue and green to red for the highest redshifts). We adopt the cut given by the dashed red line, $([3.6]-[4.5]) = 0.068([5.8]-[8.0]) - 0.075$, which effectively divides MUSIC sources into those at $z \leq 1.2$ (below the line), and at $z \geq 1.2$ (above the line).*

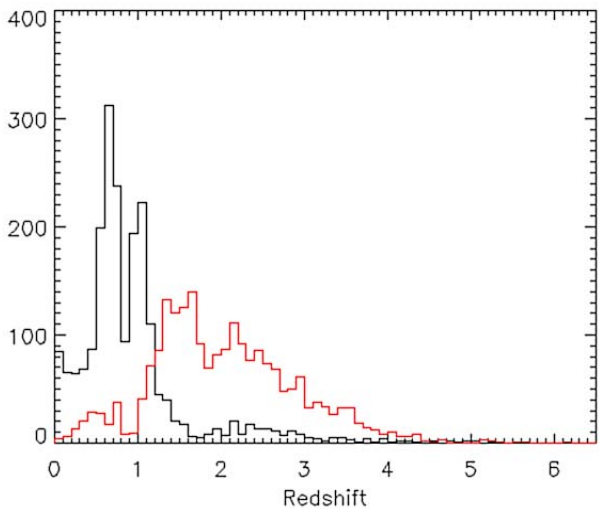

*Figure S8: Redshift histogram for MUSIC sources above (red line) and below (black line) the colour cut shown in Figure S7. There are 4242 MUSIC sources with either spectroscopic, or optical photometric redshifts. The black histogram contains 2062 objects, of which 16% are contamination from $z > 1.2$ sources. The red histogram contains 2180 objects, of which 15% are contamination from $z < 1.2$ sources.*